\documentclass{ws-mpla}

\newcommand{\nn}{\nonumber}
\newcommand{\bd}{\begin{document}}
\newcommand{\ed}{\end{document}}
\newcommand{\bc}{\begin{center}}
\newcommand{\ec}{\end{center}}
\newcommand{\be}{\begin{eqnarray}}
\newcommand{\ee}{\end{eqnarray}}
\renewcommand{\thefootnote}{\alph{footnote}}
\newcommand{\se}{\section}
\newcommand{\sse}{\subsection}
\newcommand{\text}{\rm}

\begin{document}

\markboth{C.Q.GENG AND LEI GENG}
{SOME TESTS ON CPT INVARIANCE}

%
%

\title{SOME TESTS ON CPT INVARIANCE}
 
\author{\footnotesize C.~Q. GENG\footnote{Talk presented 
at the International Conference
on {\sl Non-Perturbative Quantum Field Theory: Lattice and Beyond}.
}}

\address{
Department of Physics, National Tsing Hua University\\
Hsinchu, Taiwan, 30043\\
geng@phys.nthu.edu.tw}

\author{LEI GENG}

\address{
Department of Applied Physics\\
 University of Science and Technology Beijing\\
 Beijing, 100083}

\maketitle


\begin{abstract}
We first briefly review tests on $CPT$ invariance based on the 
consequences of the CPT theorem and then present some possible $CPT$ tests due to exotic models in which some of the CPT conditions are lost, such as
those without hermiticity.

\end{abstract}

\ccode{PACS Nos.: 11.30.Er, 14.70.Bh, 14.40.Aq}

\section{Introduction}	
Symmetry principles dictate the basic laws of physics, control structure of matter and define the fundamental forces in nature. 
The best known definition  of {\em Symmetry} can be referred to
the book by Hermann Weyl \cite{Weyl}, which stated as follows:\\

{\em ``Symmetry $\dots$ is an idea which has guided man through the centuries to the understanding and the creation of order, beauty and perfection.''}\\
\\
One can relates symmetries and conservation laws
from Noether's famous theorem that every continuous symmetry of a Lagrangian implies a conserved quantity 
and vice versa, $i.e.$,
\be\nn
Symmetries &\Longleftrightarrow& Conservation\ Laws\,.
\ee
Invariances or symmetries of translation in time, translation in space and rotation correspond to conservations of energy, momentum and angular momentum, respectively. In addition to continuous symmetries, there are discrete symmetries. In particle physics, there are three very important discrete symmetries, called
$P$, $T$ and $C$,
where 
\be
\nn  
P & : & \ parity\ or\ space\ inversion\\
\nn
&& \ \vec{x} \longleftrightarrow -\vec{x}\,,\\
\nn   
T &:&\  time\ reversal \\
\nn
&& \ t\longleftrightarrow -t\,,\\
\nn
C &:&\ particle\ (p)\ -\ antiparticle\ (\bar{p})\ exchange\\
\nn  
&& \ or\ charge\ conjugation \\
\nn
&&   \ e^-\longleftrightarrow e^+\,.
\ee   
In the following table, various physical quantities under
$P$, $T$ and $C$ transformations are given:
\be
\label{Transformation}
\matrix{
 & &P& && &T& && &C&\cr
coordinate & \vec{x} &\longrightarrow & -\vec{x} &
      & \vec{x}& \longrightarrow &\ \vec{x}
      && \vec{x}& \longrightarrow &\ \vec{x} \cr
time & t &\longrightarrow & \ t && t &\longrightarrow & -t
&& t &\longrightarrow & -t \cr
momentum & \vec{p} &\longrightarrow & -\vec{p} &
         &    \vec{p}& \longrightarrow &-\vec{p}
  &       &    \vec{p}& \longrightarrow &\ \vec{p} \cr
energy & \varepsilon &\longrightarrow &\ \varepsilon && \varepsilon &\longrightarrow & \ \varepsilon 
&& \varepsilon &\longrightarrow &\ \varepsilon \cr
angular\ momentum & \vec{J}& \longrightarrow &\ \vec{J} &
             &  \vec{J} &\longrightarrow & -\vec{J}
 &            &  \vec{J} &\longrightarrow &\ \vec{J}  \cr
spin & \vec{s} &\longrightarrow & \vec{s} &
           &    \vec{s}& \longrightarrow & -\vec{s}
  &         &    \vec{s}& \longrightarrow &\ \vec{s}  \cr
     charge & Q &\longrightarrow &\ Q
            &     &Q &\longrightarrow &\ Q
   &         &Q &\longrightarrow &-Q \cr   
electric\ field & \vec{E} &\longrightarrow &-\vec{E} &
         &  \vec{E} &\longrightarrow &\vec{E} 
    &     &  \vec{E} &\longrightarrow &- \vec{E} \cr
magnetice\ field & \vec{B} &\longrightarrow &\ \vec{B} &
           & \vec{B} &\longrightarrow &-\vec{B}
     &      & \vec{B} &\longrightarrow &-\vec{B} \cr}\\
     \nn
\ee
From the above table, we can study the symmetry properties of
$C$, $P$ and $T$ as well as their combined operations for various physical observables.  For examples,
$T$-odd product correlations of
$\vec{v}_1\cdot(\vec{v}_2\times \vec{v}_3)$ ($\vec{v}_i=\vec{s}_i$ 
or $\vec{p}_i$) 
and $\vec{s}\cdot \vec{E}$ and 
$\vec{E}\cdot\vec{B}$
are used to probe $T$ violation for K and B decays and
 electric dipole moments of particles, respectively.
 
 For a long time, physics laws were thought to be $C$, $P$
 and $T$  conserved. With these conservation laws, many experimental phenomena can be understood. For example, charge conservation law
 allows $^1S_0$ positronium atom decaying to $2\gamma$, but forbids
  $3\gamma$ mode, since $C(n\gamma)=(-1)^n$.
 Indeed, electromagnetic and strong interactions respect all these symmetries at very high accuracy. 
 However, in 1956 Lee and Yang \cite{LY} concluded that $P$ must be violated in weak interaction and in 1957 \cite{PVexpt1,PVexpt2}
 both $P$ and $C$ were found to 
 be violated maximally. To understand weak interaction, $V-A$ theory was proposed by Marshak {\em et al.} \cite{Marshak} in 1957. In this theory, 
 $P$ and $C$ are violated maximally but the combined $CP$ operation is still conserved. In 1964, CP violation (CPV) \cite{CPV}
 was observed at the level of $O(10^{-3})$ in the neutral kaon system. Recently,
 large CP violating effects have also
 been found in the neutral B decays \cite{CPVB}. Furthermore, some evidences of $T$ violation (TV) \cite{TV} was also shown in the $K^0$ system.
 It should be noted that no CPV or TV has been seen in any charged systems yet, such as those of $K^\pm$ and $B^\pm$.
Furthermore, the origin of the violations in $K^0$ and $B^0$ systems remains unclear. In the standard model, CPV or TV
arises from a unique physical phase in the Cabibbo-Kobayashi-Maskawa (CKM) quark mixing matrix \cite{ckm}. 
On the other hand, at present, there is no any experimental sign of CPT violation. 
 The experimental results can be summarized in Table 1. 

\begin{table}[h]
\tbl{Symmetry invariance and violation}
{\begin{tabular}{|c|c|c|c|c|c|}
\hline
Forces& $P$ & $C$ & $CP$ & $T$ & $CPT$ \cr
\hline
\hline
Gravity& $\surd$ & $\surd$ & $\surd$ & $\surd$ & $\surd$ \cr
\hline
Electromagnetic& $\surd$ & $\surd$ & $\surd$ & $\surd$ & $\surd$ \cr
\hline
Strong& $\surd$ & $\surd$ & $\surd$ & $\surd$ & $\surd$ \cr
\hline
Weak& $\times$ & $\times$ & $\times$ & $\times$ & $\surd$\cr
\hline
\end{tabular}}
\end{table}
\normalsize

From Table 1, it is interesting to ask that $C$, $P$, $T$
and $CP$ are broken, why not $CPT$? The immediate answer is 
that in the relativistic local quantum field theory, $CPT$ is 
invariance, so called the $CPT$ theorem \cite{CPTTh}. It holds based on three 
CPT conditions:

\begin{itemlist}
 \item Lorentz invariance,
 \item  Hermiticity of the Hamiltonian,
 \item Locality.
\end{itemlist}

\noindent
It is clear that $CPT$ is a very fundamental symmetry. If it is found to be violated, there will be tremendous impact on our fundamental physics and at least one of the three conditions above must be given up. In string theory,
 $CPT$ can be spontaneously broken. In particular, a theoretical framework with Lorentz and $CPT$ breaking terms has been formulated \cite{Kos}. The test of $CPT$ invariance is thus of considerable theoretical and experimental interest.
 
 In this talk, we will first briefly review $CPT$ tests based on the 
consequences of the CPT theorem and then study some possible $CPT$ tests due to exotic models in which one of the CPT conditions, hermiticity of the Hamiltonian, is  lost.

\section{Some Possible CPT tests }

\subsection{Consequences of the CPT theorem}
With the CPT theorem in hand, it is possible to deduce many interesting consequences. For a system with the Lagrangian ${\cal L}$ and
a local quantum field, we have 
\be
\Theta {\cal L}(\vec{x},t)\Theta^{-1} &=&{\cal L}^{\dagger}(-\vec{x},-t)
\label{L1}
\\
&=&{\cal L}(-\vec{x},-t)\,
\label{L2}
\ee
for $\Theta\equiv CPT$.
Note that Eq. (\ref{L2}) is guaranteed if ${\cal L}$ is Hermitian. 
From the table in Eq. (\ref{Transformation}), we find
\be
\Theta|p,\vec{p},\vec{s}>&=&|\bar{p},\vec{p},-\vec{s}>\,,
\ee
where $p\ (\bar{p})$ is particle (antiparticle) and
$\vec{p}$ and $\vec{s}$ are the momentum and spin.
In the following, we focus on five basic predictions 
due to the CPT theorem and we list them as tests by searching for their
possible experimental deviations. 
\\

\noindent
{\bf $\bullet$} {\em Test 1}: {\sl Mass equality between particle ($p$)
and antiparticle ($\bar{p}$)}, $i.e.$, $m_p=m_{\bar{p}}$.
\\
It is obvious that the mass of a free particle should be that of the corresponding antiparticle due to the definition of charge conjugation.
With interactions, it follows from CPT invariance even when $C$ breaks down.
One can show this by considering a particle $p$ at rest with the z-component of angular momentum $m_z$ and Hamiltonian $H$ and one has
\be
TPC|p>_{m_z}=TPe^{i\alpha_c}|\bar{p}>_{m_z}=Te^{i\alpha_c}|\bar{p}>_{m_z}
=e^{i\alpha_c}|\bar{p}>_{-m_z}
\label{E1}
\ee
where $e^{i\alpha_c}$ represents the phase factor under $C$.
 From Eq. (\ref{E1}) and $\Theta H\Theta^{-1}=H$, 
 we get
\be
m_p=<p|H|p>_{m_z}=<p|\Theta^{-1}\Theta H\Theta^{-1}\Theta|p>_{m_z}=
<\bar{p}|H|\bar{p}>_{-m_z}=m_{\bar{p}}\,.
\ee
It is clear that if there exists a difference between 
$m_p$ and $m_{\bar{p}}$ for any $p$, CPT must be violated.
Experimentally, stringent limits have been given in many systems.
In particular, one has that \cite{pdg}
\be
|m_{K^0}-m_{\bar{K}^0}|/m_{K}&<& 10^{-18}\,,
\label{exptK}
\\
|m_{e^+}-m_{e^-}|/m_e&<&8\times 10^{-9}\,.
\ee
The limit in Eq. (\ref{exptK}) is the most accurate test for the 
CPT theorem.\\

\noindent
{\bf $\bullet$} {\em Test 2}: {\sl Decay rate (lifetime) equality
 between $p$ and $\bar{p}$}, $i.e.$, $\Gamma_p=\Gamma_{\bar{p}}$
 ($\tau_p=\tau_{\bar{p}}$).\\
 We follow the proof by Bigi and Sanda \cite{BSbook}, given by
 \be\nn
 \Gamma_p&=&2\pi\sum_i\delta(m_p-\varepsilon_i)|<i;out|H_{decay}|p>|^2
 \\
 \nn
 &=&2\pi\sum_i\delta(m_p-\varepsilon_i)|<i;out|\Theta^{-1}\Theta H_{decay}\Theta^{-1}\Theta|p>|^2
\\
\nn
&=& 2\pi\sum_i\delta(m_p-\varepsilon_i)|<\bar{i};in|H_{decay}|\bar{p}>|^2
\\
&=& 2\pi\sum_i\delta(m_p-\varepsilon_i)|<\bar{i};out|H_{decay}|\bar{p}>|^2
 =\Gamma_{\bar{p}}\,,
 \label{Rate}
 \ee
 where the condition of complete sets of states has been used, $i.e.$,
 \be
 \sum_i|i;in><i;in|&=&\sum_i|i;out><i;out|=1\,.
 \ee
 Note that the decay rates in Eq. (\ref{Rate}) are for the total decay
 rates and thus $\tau_p=\tau_{\bar{p}}$ since $\tau_p=1/\Gamma_p$.
Currently, the best search is also in the $K^0$ system with \cite{pdg}
\be
(\Gamma_{K^0}-\Gamma_{\bar{K}^0})/m_{K_{average}}&=&
(7.8\pm8.4)\times 10^{-18}\,.
\ee
For the lifetime, one gets \cite{pdg}
\be
(\tau_{\mu^+}-\tau_{\mu^-})/\tau_{\mu}&=&(2\pm8)\times 10^{-5}\,.
\ee
On the other hand, if we neglect the final state interactions, we also expect the equality of partial decay rates between $p$ and $\bar{p}$.
For examples, $\Gamma(\mu^-\to e^-\bar{\nu}_e\nu_{\mu}) =
\Gamma(\mu^+\to e^+\nu_e\bar{\nu}_{\mu})$ and 
$\Gamma(K^-\to\pi^-\pi^0)=\Gamma(K^+\to\pi^+\pi^0)$.\\

 \noindent
{\bf $\bullet$} {\em Test 3}: {\sl Opposite sign of charge 
 between $p$ and $\bar{p}$}, $i.e.$, $Q_p=-Q_{\bar{p}}$.
\\ 
 From the CPT theorem, one has that $\Theta\rho(x)\Theta^{-1}=-\rho(-x)$
with $\rho(x)$ being the charge density and one obtains
\be
Q_p&=&<p|\int d^3x\rho(x)|p>_{m_z}=<p|\Theta^{-1}\Theta \int d^3x\rho(x)\Theta^{-1}\Theta|p>_{m_z}
\nn\\
&=&<\bar{p}| \int d^3x[-\rho(x)] |\bar{p}>_{-m_z}=-Q_{\bar{p}}\,.
\ee
This prediction also holds to an extremely high 
accuracy \cite{pdg}. For example, $|Q_{e^+}+Q_{e^-}|/e<4\times 10^{-8}$.\\

 \noindent
{\bf $\bullet$} {\em Test 4}: {\sl Opposite sign of 
magnetic (electric) dipole moments between
$p$ and $\bar{p}$}, 

$\ \ \ \ \ \ \  $ $i.e.$, $\mu_p=-\mu_{\bar{p}}$ 
and $d_p=-d_{\bar{p}}$.\\
For a particle $p$ in uniform and static electromagnetic fields 
$\vec{E}$ and $\vec{B}$, we find
\be
H &=& -{\mu}_p<p| \vec{s}\cdot \vec{B}|p>
 -d_p<p| \vec{s}\cdot \vec{E}|p>
 \nn\\
 &=& -{\mu}_p<p|\Theta^{-1}\Theta \vec{s}\cdot \vec{B}\Theta^{-1}\Theta|p>
 -d_p<p|\Theta^{-1}\Theta \vec{s}\cdot \vec{E}\Theta^{-1}\Theta|p>
 \nn\\
 &=& -{\mu}_p<p|\Theta^{-1}(-\vec{s}\cdot \vec{B})\Theta|p>
 -d_p<p|\Theta^{-1}(-\vec{s}\cdot \vec{E})\Theta^{-1}|p>
 \nn\\
&=&{\mu}_p<\bar{p}| \vec{s}\cdot \vec{B}|\bar{p}>
 +d_p<\bar{p}| \vec{s}\cdot \vec{E}|\bar{p}>\,,
 \label{M1}
 \\
\Theta^{-1}H\Theta&=& 
-{\mu}_{\bar{p}}<\bar{p}| \vec{s}\cdot \vec{B}|\bar{p}>
 -d_{\bar{p}}<\bar{p}|\vec{s}\cdot \vec{E}|\bar{p}>\,,
 \label{M2}
\ee
which lead to $\mu_p=-\mu_{\bar{p}}$ 
and $d_p=-d_{\bar{p}}$ based on $\Theta^{-1}H\Theta=H$.
It is interesting to note that from Eqs. (\ref{M1}) and (\ref{M2}), we conclude that both magnetic and electric dipole moments
of  Majorana neutrinos must vanish if CPT is conserved.
Experimentally, by defining $\mu_p/(e\hbar/2m_p)-1=(g_p-1)/2$, 
one has~\cite{pdg}
\be
(g_{e^+}-g_{e^-})/g_{average}&=&(-0.5\pm2.1)\times 10^{-12}\,.
\ee

\noindent
{\bf $\bullet$} {\em Test 5}: {\sl Opposite sign between transverse muon polarizations of $K^+\to\pi^0\mu^+\nu_{\mu}$ 

$\ \ \ \ \ \ \ $ and $K^-\to\pi^0\mu^-\bar{\nu}_{\mu}$,
 $i.e.$, $P_{T}({\mu^+})=-P_{T}({\mu^-})$.}\\
 The transverse muon polarization is associated with 
 $\vec{s}_\mu\cdot(\vec{p}_{\pi}\times \vec{p}_\mu)$ 
 which itself is a $P$-even and $T$-odd quantity. 
 $P_T(\mu^+)$ has been used to search for $CP$ violation induced by new physics \cite{BG} since it is vanishingly small in the standard model. The current experimental data is $P_T(\mu^+)=(-1.7\pm2.3\pm1.1)\times 10^{-3}$, given by the E246 collaboration at KEK \cite{E246}.
 
   The consequence of the CPT theorem in this test can be easily  
   proved as follows:
 \be
 &&P_T(\mu^+)<\Theta^{-1}\Theta\vec{s}_{\mu^+}\cdot(\vec{p}_{\pi}\times \vec{p}_{\mu^+})\Theta^{-1}\Theta>_{\mu^+}
 \nn\\
 &=&P_T(\mu^+)<\Theta^{-1}[-\vec{s}_{\mu^-}\cdot(\vec{p}_{\pi}\times \vec{p}_{\mu^-})]\Theta>_{\mu^+}
 \nn\\
 &=&-P_T(\mu^+)<\vec{s}_{\mu^-}\cdot(\vec{p}_{\pi}\times \vec{p}_{\mu^-})>_{\mu^-}
 \nn\\
 &=&
P_T(\mu^-)<\vec{s}_{\mu^-}\cdot(\vec{p}_{\pi}\times \vec{p}_{\mu^-})>_{\mu^-}\,.
\ee
Note that final state interactions give the same amount of contributions
to $P_T(\mu^-)$ and $P_T(\mu^-)$ but very small $O(10^{-6})$.

\subsection{Models without hermiticity}
We now examine models without hermiticity as first suggested by Okun \cite{Okun}.
In these models, Eq. (\ref{L2}) doesn't hold anymore and CPT could be broken. Moreover, the breaking down of hermiticity also breaks unitarity of $S$-matrix. We note that one cannot exclude this possibility from appearing in future microscopic quantum gravity as emphasized by Mavromatos \cite{QG}.
In this talk, we only concentrate on the CPT violating effects.
In the following,
we will present some examples as possible $CPT$ tests due to this exotic 
scenario.\\

\noindent
{\bf $\bullet$} {\em Test 1}: 
{\sl Circular photon polarizations in $\Phi\to\gamma\gamma\ (\Phi=\pi^0,\eta)$} \cite{Okun}\\
The most general effective interaction for $\Phi\to\gamma\gamma$ can be described by 
\be
{\cal L}&=&
g\Phi F_{\mu\nu}F_{\alpha\beta}\varepsilon^{\mu\nu\alpha\beta}\;+\;
h\Phi F_{\mu\nu}F^{\mu\nu}
\nn\\
&=&
{\cal L}_p\;+\;{\cal L}_s\,.
\label{Lsp}
\ee
The term of ${\cal L}_p$ in Eq. (\ref{Lsp}) is the normal anomaly term which gives  to
the decay rates.
In Table \ref{Lcpt}, 
\begin{table}[h]
\tbl{C, P and T transformations for $\Phi$, $F_{\mu\nu}F_{\alpha\beta}\varepsilon^{\mu\nu\alpha\beta}$
and $F_{\mu\nu}F^{\mu\nu}$}
{\begin{tabular}{|c|c|c|c|c|c|}
\hline
Quantity& $P$ & $C$ & $T$ & $CP$ & $CPT$ \cr
\hline
\hline
$\Phi$& $-$ & $+$ & $-$ & $-$ & $+$ \cr
\hline
$F_{\mu\nu}F_{\alpha\beta}\varepsilon^{\mu\nu\alpha\beta}$
& $-$ & $+$ & $-$ & $-$ & $+$ \cr
\hline
$F_{\mu\nu}F^{\mu\nu}$& $+$ & $+$ & $+$ & $+$ & $+$ \cr
\hline
\end{tabular}}
\label{Lcpt}
\end{table}
\normalsize
we show the properties of C, P and T  
for $\Phi$, $F_{\mu\nu}F_{\alpha\beta}\varepsilon^{\mu\nu\alpha\beta}$
and $F_{\mu\nu}F^{\mu\nu}$.
Note that
\be
 F_{\mu\nu}F_{\alpha\beta}\varepsilon^{\mu\nu\alpha\beta}
\propto \vec{E}\cdot\vec{B}\,,\ \
F_{\mu\nu}F^{\mu\nu}\propto E^2-B^2\,.
\ee
From Table \ref{Lcpt}, it is easy to see that the
{\em hermiticity} of the Lagrangian in Eq. (\ref{Lsp}) requires
that both $g$ and $h$ must be real numbers, $i.e.$, $g=Re g$ and $h=Re h$.
In this case, CPT is a good symmetry for both ${\cal L}_s$ and ${\cal L}_p$.
In general, Both $g$ and $h$ can be complex if hermiticity is broken.
From the Lagrangian in Eq. (\ref{Lsp}), by summing over one of the photon polarization, we find that the squared amplitude is given by
\be
|{\cal M}|^2&\propto& (4|g|^2+|h|^2)[q\cdot k (q^\mu k^\nu+q^\nu k^\mu)-(q\cdot k)^2g^{\mu\nu}]\varepsilon_\mu\varepsilon_\nu^*
\nn\\
&&+4i Im(gh^*)(q\cdot k)\varepsilon^{\mu\nu\alpha\beta}k_\alpha q_\beta
\varepsilon_\mu\varepsilon_\nu^*\,,
\label{Ampl}
\ee
where $q$ and $k$ are the 4-momenta of the two photon and $\varepsilon$
is one of the photon 4-polarization vector.
It is interesting to observe that in the rest frame of $\Phi$
the 2nd term in Eq. (\ref{Ampl})
is related to
\be
(\vec{\varepsilon}_1\times \vec{\varepsilon}_2)\cdot \vec{k}\,,
\label{Pol}
\ee
which is the so called circular photon polarization.
It is clear that with hermiticity no circular $\gamma$ polarization
can be induced due to $Im(gh^*)=0$.
We now consider two extreme cases without hermiticity in Eq. (\ref{Lsp}),
which are shown in Table \ref{Cases}. 
\begin{table}[h]
\tbl{C, P and T transformations for ${\cal L}_p$ and ${\cal L}_s$ in the two cases.}
{\begin{tabular}{|c|c|c|}
\hline
& Case 1: & Case 2: \cr
\hline
&$g=iIm\, g\ (Re\, g=0)$ & $g=Re\, g\ (Im\, g=0)$ \cr
&$h=Re\, h\ (Im\, h=0)$ & $h=iIm\, h\ (Re\, h=0)$ \cr
\hline
& $P$~~~$C$~~~$T$~~~$CP$~~~$CPT$ &$P$~~~$C$~~~$T$~~~$CP$~~~$CPT$\cr
\hline
${\cal L}_p$&~+~~~~+~~~~--~~~~~+~~~~~~--~~~~~&+~~~+~~~+~~~~+~~~~~+~~~~ \cr
\hline
${\cal L}_s$&~+~~~~--~~~~~--~~~~~--~~~~~~~+~~~~&~+~~~~--~~~+~~~~--~~~~~~--~~~~~ \cr
\hline
\end{tabular}}
\label{Cases}
\end{table}
\normalsize
In both cases, CPT is broken and
a nonzero circular $\gamma$ polarzation is expected.
It is interesting to note that imaginary parts of $g$ and $h$ can induce
imaginary values of proton magnetic and electric dipole moments \cite{Okun}, respectively, which of course are CPT violating quantities.
Finally, we remark that the circular $\gamma$ polarzation
can be also generated by the final state interaction such as the two-photon rescattering though electron box diagram \cite{Okun}. In this case, there is no CPT violation since $Im(gh^*)=0$. \\

\noindent
{\bf $\bullet$} {\em Test 2}: 
{\sl Circular photon polarization in ${^1S}_0\,(e^+e^-)\to\gamma\gamma$}\\
In Eqs. (\ref{M1}) and (\ref{M2}), we have assumed that $d_p$
is real. If CPT is broken, $d_p$ could be complex. 
If electron  has an electric dipole moment, $d_e$, one can write the effective interaction as
\be
{\cal L}_{edm}&=& -{i\over 2}d_e\bar{e}\sigma_{\mu\nu}\gamma_5eF^{\mu\nu}\,,
\label{Ledm}
\ee
induced by loops. Note that hermiticity requires that $d_e$ must be real.
Now, beside the tree QED diagram contribution,
  one gets contributions to 
$e^+e^-\to\gamma\gamma$ with replacing one vertex ($\gamma_\mu$)
 in the tree diagram
by the edm vertex in Eq. (\ref{Ledm}). The interferences between these two types of contributions lead to a non-zero value of circular $\gamma$ polarization \cite{Okun} in Eq. (\ref{Pol}) if $Im (d_e)\neq 0$. Clearly, CPT must be broken.\\

\noindent
{\bf $\bullet$} {\em Test 3}: 
{\sl Circular photon polarization in $\eta\to\pi^+\pi^-\gamma$}\\
In Refs.~\refcite{gengeta1} and \refcite{gengeta2}, we have studied CP violation
in $\eta\to\pi^+\pi^-\gamma$ with the CPT theorem. Here we shall discuss it without imposing the $CPT$ symmetry. In the $\eta$ rest frame,
The most general decay amplitude is given by 
\be
{\cal M} &=& im_{\eta}^{-2}E_{\gamma}[M\hat{k}\cdot (\vec{\varepsilon}\times 
\vec{p}_+)-E\vec{\varepsilon}\cdot\vec{p}_+]\,,
\label{Amp2}
\ee	
where $\vec{p}_{\pm}$ and  $\vec{k}$, $E_\gamma$ and $\vec{\varepsilon}$
are momenta of $\pi^{\pm}$ and 
the phone momentum, energy and polarization, respectively. In the absence of final state interactions, $M$ and $E$ are purely real if the CPT theorem holds according to the table in Eq. (\ref{Transformation}).
The squared amplitude from Eq. (\ref{Amp2}) is given by
\be
|{\cal M}|^2 &=&
m_{\eta}^{-4}E_{\gamma}^2\left\{ 
|M|^2|\hat{k}\cdot (\vec{\epsilon}\times 
\vec{p}_+)|^2+|E|^2|\vec{\epsilon}\cdot\vec{p}_+|^2+
\right.
\nn\\
&&
\left.
 E^*M[\hat{k}\cdot (\vec{p}_+\times\vec{\epsilon})]
\,(\vec{\epsilon}\cdot\vec{p}_+)^*+
M^*E[\hat{k}\cdot (\vec{p}_+\times\vec{\epsilon})]^*
\,(\vec{\epsilon}\cdot\vec{p}_+)\right\}\,.
\label{Amp3}
\ee
It is easy to show \cite{gengeta1,gengeta2} that the circular photon polarization is found to be
\be
S_2(E_\gamma) & \propto & 2 Im \left( E^*M \right) / \left( |E|^2 + |M|^2 
\right)\,,
\label{S2}
\ee
which is zero if both $E$ and $M$ are real.
Phenomenologically, the decay rate of $\eta\to\pi^+\pi^-\gamma$ is 
described by a real $M$ term from the box-anomaly and resonance 
contributions \cite{Eta}.
Possible interactions which could yield the $E$ 
term in Eq. (\ref{Amp2}) were studied in Refs.~\refcite{gengeta1,gengeta2}.
In particular,
a four-fermion operator was introduced, 
given by
\be
{\cal O} &=& {1\over m_{\eta}^3}G\,\bar{s}
i\sigma_{\mu\nu}\gamma_5(p-k)^{\nu}\,s\,
\bar{u}\gamma^{\mu}u\,,
\label{O}
\ee
where $u (s)$ stands for the up (strange) quark and
$G$ is a dimensionless parameter originating from 
yet unknown 
short distance 
physics. The interaction in Eq. (\ref{O})
leads to 
\be
E&\sim & 2eF(E_\gamma)G\,,
\label{Ei}
\ee
where $F(E_\gamma)$ is a real form factor which is a function of $E_\gamma$.
If $G$ contains an imaginary part, the 
circular $\gamma$ polarization $S_2$ in Eq. (\ref{S2}) is nonzero and CPT is broken. Note that it is possible that Eq. (\ref{O}) can be generated by the
electric dipole moment of the strange quark.

\section{Summary and Remarks}
There is 
no evidence of CPT Violation so far.
The CPT violating effects such as the mass inequality between
particle and antiparticle
are found and expected to be vanishingly small due to the CPT theorem.
There exist some exotic models which break hermiticity in the Lagrangians,
in which CPT is also broken. Some consequences in these exotic models have been discussed. In particular, the circular photon polarizations have been studied. Finally, we remark that we have not examined CPT violating effects due to other CPT violating theories with hermiticity, in particular the
recent ones based on
the Lorentz and CPT violating terms in the
modified Dirac equation \cite{Kos}.

\section*{Acknowledgments}

This work was supported in part by
 the National Science Council of the Republic of China under
 Contract \#:  NSC-93-2112-M-007-014.

\end{document}